\documentclass[11pt]{article}
\usepackage{setspace}
\usepackage{amssymb}
\usepackage{amsmath}
\usepackage{amsfonts}

\def\be{\begin{equation}}

\def\ee{\end{equation}}

\def\dd{\partial}

\newcommand\of[1]{\left( #1 \right)}

\def\bea{\begin{eqnarray}}

\def\eea{\end{eqnarray}}

\def\half{\frac{1}{2}}

\makeatletter
\def\blfootnote{\xdef\@thefnmark{}\@footnotetext}
\makeatother

\begin{document}

\singlespace

\begin{flushright} BRX TH-6288 \\
CALT-TH 2015-001
\end{flushright}

\vspace*{.3in}

\begin{center}

{\Large\bf Bel-Robinson as stress-tensor gradients and their extensions to massive spin ($0$,$1$,$2$)}

{\large S.\ Deser}

{\it 
Walter Burke Institute for Theoretical Physics, \\
California Institute of Technology, Pasadena, CA 91125; \\
Physics Department,  Brandeis University, Waltham, MA 02454 \\
{\tt deser@brandeis.edu}
}

{\large J.\ Franklin}

{\it Physics Department, Reed College, Portland, OR 97202 \\
{\tt jfrankli@reed.edu}}

\end{center}

\begin{abstract}
We show that the Bel-Robinson (BR) tensor is -- generically, as well as in its original GR setting -- an autonomously conserved  part of the, manifestly conserved, double gradient of a system's stress-tensor. This suggests its natural extension from GR  to matter models, first to (known) massless scalars and vectors, then to massive ones, including tensors. These massive versions
are to be expected, given that they arise upon KK reduction of massless $D+1$ ones. We exhibit the resulting spin $(0, 1, 2)$ ``massive" BR.
\end{abstract}

\section{Introduction}
The motivation for the present Bel-Robinson (BR) [1] exercise is twofold: First, to dispel any lingering mystery about these conserved four-index tensors' existence by showing that one of their avatars is just a -- separately conserved -- part of a stress-tensor's double gradient, exactly as is the case in GR [2]. Second, to exploit this fact by extending BR first to massless, then to massive matter. Our explicit examples will include scalars, vectors (Proca) and massive tensor -- Fierz-Pauli (FP) -- spin 2, as well as a ``how-to" outline for nonlinear -- massive gravity's (mGR) BR. The existence of massive BR extensions is separately suggested by the fact that massless systems in $D+1$ become massive in $D$ under a periodic KK compactification of the ``fifth" dimension [3], even though, as we will see, the $D=5$ BR neither does nor should -- unlike the action or the field equations -- reduce directly to the massive $D=4$ form.

\section{BR-stress tensor connection}

Consider for concreteness the free, initially $m=0$, scalar field in any $D$, whose on-shell ($\Box \phi=0$) conserved $T_{\mu\nu}$ is
\begin{equation}
T_{\mu\nu} = \phi_{,\mu} \, \phi_{,\nu} - \frac{1}{2} \, \eta_{\mu\nu} \, \phi_{,\gamma} \, \phi_{,}^\gamma \, \, ;
\end{equation}
obviously $\dd_\alpha \, \dd_\beta \, T_{\mu\nu}$ remains conserved on ($\mu\nu$). Slightly less obvious, but fundamental to our aim, is the separate conservation of its ``spread-out" derivatives and the remainder, defined by
\begin{equation}
\begin{aligned}
\dd_\alpha \, \dd_\beta \, T_{\mu\nu} &= \phi_{,\mu\beta} \, \phi_{,\nu\alpha} +  \phi_{,\mu\alpha} \, \phi_{,\nu\beta} - \eta_{\mu\nu} \, \phi_{,\gamma\beta} \, \phi^{\, \gamma}_{,\, \, \, \alpha} 
+ \left[ \phi_{,\mu\beta\alpha} \, \phi_{,\nu} + \phi_{,\mu} \, \phi_{,\nu\beta\alpha} - \eta_{\mu\nu} \, \phi_{,\gamma \beta \alpha} \, \phi_{,}^{\, \gamma}\right] \\
&\equiv B^0_{\mu\nu\alpha\beta} + \left[\Delta_{\mu\nu\alpha\beta}\right]
\end{aligned}
\end{equation}
Their independent conservation is trivially verified explicitly, or by noting that $\Delta$'s higher derivative structure requires it. We will see that the same holds for higher spins, so henceforth we drop $\Delta$, and concentrate (mostly) on  the ``spread-out" parts of $\dd \dd T$.  

The extension to massive scalars follows the same prescription, namely
adding the ``spread-out" $\dd_\alpha \, \dd_\beta$ acting on the mass part of $T_{\mu\nu}$, here $-1/2 \, m^2 \, \eta_{\mu\nu} \phi^2$  in our mostly plus signature, where 
$(\Box - m^2) \, \phi=0$. The compactification of $\Box_{(D+1)}$ with periodic 
$m$-dependence of the ``5th" dimension is of course precisely $(\Box_D -m^2)$. For completeness, we display the massive $B^m_{\mu\nu\alpha\beta}$:
\begin{equation}
B^m_{\mu\nu\alpha\beta} = \phi_{,\mu\beta}\, \phi_{,\nu\alpha} +  \phi_{,\mu\alpha} \, \phi_{,\nu\beta}  - \eta_{\mu\nu} \, \phi_{,\gamma\beta} \, \phi^{\, \gamma}_{,\, \, \, \alpha} 
-m^2 \, \eta_{\mu\nu} \, \phi_{,\alpha} \, \phi_{,\beta}
\end{equation}
We may now explain the disparity mentioned above between the 
$D$-components of $\ ^{(D+1)} B^0_{\mu\nu\alpha\beta}$ and the massive  $B^m_{\mu\nu\alpha\beta}$ in $D$.
That they should not agree is obvious from the fact that their conservation properties differ, since only
\begin{equation}
\dd^{\mu} \ ^{(D+1)} B^0_{\mu\nu\alpha\beta} +\dd^5 \, \ ^{(D+1)} B^0_{5\nu\alpha\beta} =0,   
\end{equation}
and of course the second term does not vanish. It is easy to verify that
this ``discrepancy" between these two BR is precisely what is required for their respective conservations.

\section{BR for $s = 1$, Proca fields}
The higher spin equivalents of our scalar system follow exactly the latter's pattern above. For vectors, start with $T_{\mu\nu} = F_{\mu\alpha}\, F^{\, \, \,\alpha}_{\nu} -1/4\, \eta_{\mu\nu}\, F_{\alpha\beta} \, F^{\alpha\beta}$, apply the $\dd_\alpha \, \dd_\beta$ prescription, and add the $\dd_\alpha \, \dd_\beta$ of $\Delta T_{\mu\nu} =m^2 \,(A_\mu\, A_\nu -1/2\, \eta_{\mu\nu}\,  A_\alpha \, A^\alpha)$. The conserved (on $(\mu\nu)$ only) form of the Proca BR is thus
\begin{equation}
B^m_{\mu\nu\alpha\beta}(A_\mu) = F_{\mu\sigma,\alpha} \, F_{\nu \, \, ,\beta}^{\, \, \sigma} + F_{\mu\sigma,\beta} \, F_{\nu \, \, ,\alpha}^{\, \, \sigma} - \frac{1}{2} \, \eta_{\mu\nu} \, F^{\sigma\tau}_{\, \, \, \, \, \,  ,\alpha} \, F_{\sigma\tau,\beta} 
+ m^2 \, \of{ A_{\mu,\alpha} \, A_{\nu,\beta} + A_{\mu,\beta} \, A_{\nu,\alpha} - \eta_{\mu\nu} \, A_{\sigma,\beta} \, A^\sigma_{\, \, ,\alpha}}.
\end{equation}
For $m=0$, it agrees with the original Maxwell version [4]. Conservation of B in all cases (including GR) consists of two separate parts: one uses the ``Bianchi" identities, here the cyclic 
$\dd_{[\mu} \, F_{\nu\sigma]} = 0$, while the other depends on the dynamics, here $\dd^\mu \, F_{\mu\nu} - m^2 \, A_\nu= 0$, and similarly for the scalars
treated above. The mass terms only affect the dynamical part, of course.

\section{Spin 2}
Our new example is the massive spin 2 -- Fierz-Pauli (FP) field, consisting of linearized GR as kinetic, plus a -- unique -- mass, term. This model has received enormous recent attention as the starting-point for so-called massive gravity (mGR), its nonlinear extension with full GR plus a non-derivative mass term, necessarily involving a fixed background metric.     
Rather than proceeding via the (linearized) BR route, it is instructive to first start with the FP stress-tensor, as for lower spins. The latter, unlike that of linearized GR alone, is perfectly well-defined, there no longer being any gauge invariance to respect. We will use a -- legitimate -- shortcut below, namely since we are only interested in $T_{\mu\nu}$ and its related $\dd \dd T$ ``BR" forms on-shell -- they are not dynamical currents -- we apply the five constraints  $\dd^\mu h_{\mu\nu}=0= h_\mu^\mu$ on the $h_{\mu\nu}$ field that reduce it to describe just the $2s+1=5$ spin $2$ excitations. As a result, the action is simply
\begin{equation}
I(h)= -1/2 \int d^4x [(h_{\mu\nu,\alpha})^2 +m^2 \, (h_{\mu\nu})^2]; 
\end{equation}
here all indices are moved by the -- mostly plus -- Minkowski metric $\eta_{\mu\nu}$, so the field equations are $(\Box-m^2) h_{\mu\nu}=0$. 
To obtain the stress tensor, we could follow the Belinfante procedure: turn all $\eta_{\mu\nu} \rightarrow g_{\mu\nu}(x)$, vary the action with respect to $g_{\mu\nu}$, then set them back to $\eta_{\mu\nu}$ to define $T_{\mu\nu}$.  In addition to the $g^{\mu\nu}$ and the overall $\sqrt{-g}$ factor in (6), a new (higher spin) phenomenon arises because the partial derivatives  $\dd_\alpha$ now become covariant ones, $D_\alpha = (\dd_\alpha - \Gamma)$ acting on the tensors $h_{\mu\nu}$. Their variations, 
$\Delta \Gamma \sim \dd (\Delta g)$ give rise to novel contributions to $T_{\mu\nu}$, namely  $\sim \dd (h \dd h)$, that is not only $\sim \dd h \dd h$, but also $\sim h\, \dd \dd h$, absent in the lower spins. A dull calculation yields
\begin{equation}
\begin{aligned}
T_{\mu\nu} &=T(g)_{\mu\nu} + T(\Gamma)_{\mu\nu} \\
T(g)_{\mu\nu} &=2 h_{\mu\alpha,\beta} h_{\nu \alpha,\beta} +h_{\alpha\beta,\mu}\, h_{\alpha\beta,\nu} -\frac{1}{2} \, \eta_{\mu\nu} (\dd_\alpha \, h_{\beta\gamma})^2 
+ m^2\, \left[2 \, h_{\mu\alpha} h_{\nu\alpha}-\frac{1}{2} \eta_{\mu\nu} (h_{\gamma\delta})^2\right] \\
T(\Gamma)_{\mu\nu} &=h_{\beta\gamma} (h_{\mu\gamma,\nu\beta} + \nu \leftrightarrow \mu) - (h_{\nu\gamma,\beta} h_{\beta\gamma,\mu}  + \nu \leftrightarrow \mu) 
-2 h_{\nu\gamma,\beta} h_{\mu\gamma,\beta} - (h_{\nu\gamma} \Box h_{\mu\gamma}  + \mu \leftrightarrow \nu)
\end{aligned}
\end{equation}
which is easily verified to be conserved on-shell, namely with 
$(\Box-m^2) h_{\mu\nu}=0= \dd_\mu \, h^{\mu\nu} =0= h^\alpha_\alpha$. This $T_{\mu\nu}$ is of course not unique: there is always an ambiguity -- for all stress-tensors -- of adding identically conserved ``superpotentials" $\sim \dd_\beta \dd_\delta H^{[\mu\beta] [\nu\delta]}$, that also do not contribute to the integrated  $(P_\mu , J_{\mu\nu})$ (and of asymmetric 
$\dd_\alpha H^{[\alpha\mu]\nu}$ of the type present in the Einstein pseudo-tensor). Since we are concerned with the ``spread" $\dd_\alpha \dd_\beta$ of $T_{\mu\nu}$, the $h\, \dd\dd h$ terms present a potential obstacle. Instead of removing them by finding a suitable superpotential, we use the already available Landau-Lifshits (LL) pseudotensor, since it embodies all the needed properties, being symmetric, conserved on-shell (also for $m\ne 0$, as can easily be checked from the detailed form below) and -- most important -- of the desired $\dd h \, \dd h$ form. Including the mass contribution, and using $h_{\mu\nu}$'s traceless-divergencelessness, it reduces to:
\begin{equation}
\begin{aligned}
T^m_{\mu\nu}(\hbox{LL}) &= \half \, \eta_{\mu\nu} \, h_{\beta \sigma,\rho} \, h_{\beta\rho,\sigma} - h_{\nu\beta,\rho} \, h_{\beta\rho,\mu} - h_{\mu \beta,\rho} \, h_{\beta\rho,\nu} 
+ h_{\mu\beta,\rho} \, h_{\nu\beta,\rho} + \half \, h_{\rho\lambda,\mu} \, h_{\rho\lambda,\nu} - \frac{1}{4} \, \eta_{\mu\nu} (h_{\rho\lambda,\beta})^2\\
&+ m^2 \, [h_{\mu\alpha} \, h_{\nu\alpha} - \frac{1}{4} \, \eta_{\mu\nu} \, h_{\alpha\beta}^2] = T^m_{\nu\mu}(\hbox{LL}).
\end{aligned}
\end{equation}
Now we may safely proceed to take its ``spread", $B^m_{\mu\nu\alpha\beta} \dot= \dd_\alpha \dd_\beta\, T^m_{\mu\nu}(\hbox{LL})$, $4$-index extension as for lower spins. We need not spell out the obvious result, 
$B^m_{\mu\nu\alpha\beta} \sim  \dd_\alpha \dd_\beta T^{m}_{\mu\nu}(\hbox{LL}) \sim (\dd \dd h)\, (\dd \dd h) +m^2 \, (\dd h \,  \dd h)$, with the assurance that it is indeed conserved on FP shell (as is easily checked as well).

The alternate route to the above ``low spin" procedure is to return to FP's GR roots and use a true (linearized, of course) BR kinetic term, namely:
\begin{equation}
B^0_{\mu\nu\alpha\beta} = R_{\sigma\mu\tau\alpha} \, R_{\sigma\nu\tau\beta} + R_{\sigma\mu\tau\beta} \, R_{\sigma\nu\tau\alpha} - \frac{1}{8} \, g_{\mu\nu} \, g_{\alpha\beta} \, R^{\lambda\sigma\tau\delta} \, R_{\lambda\sigma\tau\delta},
\end{equation}
whose divergence is
\begin{equation}\label{divB}
B^0_{\mu\nu\alpha\beta,\mu} = R_{\sigma\nu\tau\beta} \, \left(R_{\sigma \tau,\alpha} - R_{\sigma \alpha,\tau} \right) + \alpha \leftrightarrow \beta
\end{equation}
using the Bianchi identity.  [This form of BR is equivalent to 
\begin{equation}
B^0_{\mu\nu\alpha\beta} = R_{\sigma\mu\tau\alpha} \, R_{\sigma\nu\tau\beta} + R_{\sigma\mu\tau\beta} \, R_{\sigma\nu\tau\alpha} - \frac{1}{2} \, g_{\mu\nu} \, R_{\alpha \sigma\tau\delta} \, R_{\beta\sigma\tau\delta},
\end{equation}
(only) in $D=4$ where $R_{\mu\alpha\beta\gamma} \, R_{\nu\alpha\beta\gamma} = \frac{1}{4} \, g_{\mu\nu} \, R_{\alpha\beta\gamma\delta} \, R_{\alpha\beta\gamma\delta}$].  On FP shell, 
\begin{equation}
R_{\mu\nu\alpha\beta,\mu} = -\frac{1}{2} \, m^2 \, \left( h_{\beta\nu,\alpha} - h_{\alpha\nu,\beta} \right) \, \, \, \, \, \, \, \, \, \, \, 
R_{\nu\beta} = -\frac{1}{2} \, m^2 \, h_{\beta\nu},
\end{equation}
we get
\begin{equation}
B^0_{\mu\nu\alpha\beta,\mu} = -\frac{1}{2} \, m^2 \, R_{\sigma\nu\tau\beta} \, \left( h_{\sigma\tau,\alpha} - h_{\sigma\alpha,\tau} \right) + \alpha \leftrightarrow \beta.
\end{equation}
Now we add to $B^0$ of (9) or (11) the mass term 
\begin{equation}
\begin{aligned}
\Delta B_{\mu\nu\alpha\beta} &= m^2 \, \biggl[ \eta_{\mu\nu} \, \of{h_{\alpha\sigma,\tau} \, h_{\beta\sigma,\tau} -  h_{\sigma\tau,\alpha} \, h_{\beta\sigma,\tau}  - h_{\sigma\tau,\beta} \, h_{\alpha\sigma,\tau} } \\
&-\of{ h_{\alpha\mu,\sigma} \, h_{\beta\nu,\sigma} - h_{\alpha\mu,\sigma} \, h_{\sigma\nu,\beta} - h_{\mu\sigma,\alpha} \, h_{\beta\nu,\sigma} + h_{\mu\sigma,\alpha} \, h_{\nu\sigma,\beta} + \alpha \leftrightarrow \beta} 
\biggr].
\end{aligned}
\end{equation}
It is then easy to prove that $B^m \dot = B^0 + \Delta B$ is conserved on $(\mu\nu)$.

The second massive model is, as we mentioned, massive gravity (mGR); it adds to the full GR action a $3$-parameter family of mass terms that depend on both the full ``metric" $g_{\mu\nu}$ of the Einstein part and on a fixed background $f_{\mu\nu}$. This hybrid is NOT of course a Riemann space theory but a (complicated) spin-$2$ field action in a fixed, say flat, space $f_{\mu\nu}$, where  local stress tensors are well-defined. What changes from (9,11), then, is that B$^{\hbox{\tiny GR}}$ is now the full nonlinear one, as is the $\Delta T \sim m^2$ contribution. The dependence on $f_{\mu\nu}$ being extremely complicated for all three allowed (by ghost-freedom) mass terms, we have not bothered to compute their $\Delta T_{\mu\nu}$ -- they are sufficiently sick [5] to warrant omission. Nor have we considered the last variant in which both ($g_{\mu\nu}$ and $f_{\mu\nu}$) are dynamical, their action being the sum of their full Einstein contributions plus a non-derivative coupling term $\sim m^2 M(f,g)$. Again, since this model is likely to share the acausality of mGR, we have not looked for its BR.  These formal tasks we are happy to leave open.

\section{Summary}

We have demystified and generalized the venerable BR tensor beyond its GR roots, both by expressing its (known) avatars for massless scalar and vector matter fields as conserved double gradient parts of their stress tensors, and by generalizing it further to include their massive extensions for spins $(0,1)$, as supported by KK considerations. For spin $2$, we exhibited free massive spin $2$ FP's BR and noted how it could be obtained for nonlinear massive gravity models. Our process was greatly simplified throughout by the separate conservation of the two disparate parts of $\dd \dd \, T_{\mu\nu}$, allowing us to concentrate on the more physical, ``spread derivatives" component.

Finally, while we have worked in flat space for simplicity, suitable covariantizing of the massless scalar and vector BR is known.The resulting total BR -- with extra non-minimal terms required by non-commutativity of covariant derivatives -- are now (covariantly) conserved on combined Einstein+matter shell. The mass terms' contributions are now of course to be added as $\sim D\, D\, \Delta T_{\mu\nu}$. We have not carried out FP's (or even linearized spin $2$) BR covariantization, but it may present
a new set of problems, since at least massless tensor fields cannot consistently couple [6] to GR, except of course when $h_{\mu\nu}$ is the linearized metric deviation about the background!

\subsection*{Acknowledgements}
SD was supported in part by Grants NSF PHY- 1266107 and DOE \# DE-SC0011632.

\end{document}